\documentclass[final,3p,times,twocolumn]{elsarticle}
\usepackage{amssymb}
\usepackage{amsmath}
\def \be{\begin{equation}}
\def \ee{\end{equation}} 

\journal{Physica E}

\begin{document}

\begin{frontmatter}

%% Title, authors and addresses

%% use the tnoteref command within \title for footnotes;
%% use the tnotetext command for theassociated footnote;
%% use the fnref command within \author or \address for footnotes;
%% use the fntext command for theassociated footnote;
%% use the corref command within \author for corresponding author footnotes;
%% use the cortext command for theassociated footnote;
%% use the ead command for the email address,
%% and the form \ead[url] for the home page:
%% \title{Title\tnoteref{label1}}
%% \tnotetext[label1]{}
%% \author{Name\corref{cor1}\fnref{label2}}
%% \ead{email address}
%% \ead[url]{home page}
%% \fntext[label2]{}
%% \cortext[cor1]{}
%% \address{Address\fnref{label3}}
%% \fntext[label3]{}

\title{Absorbing/Emitting Phonons with one dimensional MOSFETs}

%% use optional labels to link authors explicitly to addresses:
%% \author[label1,label2]{}
%% \address[label1]{}

%% \address[label2]{}

\author{Riccardo Bosisio\fnref{label2}}
\ead{riccardo.bosisio@nano.cnr.it}
\author{Cosimo Gorini\fnref{label3}}
\author{Genevi\`eve Fleury}
\author{Jean-Louis Pichard} 
\address{Service de Physique de l'\'Etat Condens\'e, (CNRS UMR 3680), DSM/IRAMIS/SPEC, CEA Saclay, 91191 Gif-sur-Yvette, France}
\fntext[label2]{Present address: NEST, Istituto Nanoscienze, Piazza San Silvestro 12, 56127 Pisa, Italy} 
\fntext[label3]{Present Address: Institut fur Theoretische Physik, Universitat Regensburg, 93040 Regensburg, Germany} 

%\author{label1,label3]{Cosimo Gorini} 
%\author[label1]{Genevi\`eve Fleury}
%\author[label1]{Jean-Louis Pichard}
%\address[label1]{Service de Physique de l'\'Etat Condens\'e, (CNRS UMR 3680), DSM/IRAMIS/SPEC, CEA Saclay, 91191 Gif-sur-Yvette, France}
%\address[label2]{NEST, Istituto Nanoscienze, Piazza San Silvestro 12, 56127 Pisa, Italy} 
%\address[label3]{Institut f\¨ur Theoretische Physik, Universit\¨at Regensburg, 93040 Regensburg, Germany} 

\begin{abstract}
We consider nanowires in the field effect transistor device configuration. Modeling each nanowire as a one dimensional lattice with random site 
potentials, we study the heat exchanges between the nanowire electrons and the substrate phonons, when electron transport is due to phonon-assisted 
hops between localized states. Shifting the nanowire conduction band with a metallic gate induces different behaviors.
When the Fermi potential is located near the band center, a bias voltage gives rise to small local heat exchanges which fluctuate 
randomly along the nanowire. When it is located  near one of the band edges, the bias voltage yields heat currents which flow mainly 
from the substrate towards the nanowire near one boundary of the nanowire, and in the opposite direction near the other boundary. 
This opens interesting perspectives for heat management at submicron scales: Arrays of parallel gated nanowires could be used 
for a field control of phonon emission/absorption. 
%, taking advantage of a strong energy dependence of the nanowire density of states and localization lengths. 
\end{abstract}

\begin{keyword}
%% keywords here, in the form: keyword \sep keyword
MOSFET \sep heat management \sep hopping transport \sep semiconductor nanowires

%% PACS codes here, in the form: \PACS code \sep code
\PACS 72.15.Rn \sep 72.20.Ee \sep 81.07.Gf \sep 85.30.Tv \sep 63.22.Gh 
%% MSC codes here, in the form: \MSC code \sep code
%% or \MSC[2008] code \sep code (2000 is the default)

\end{keyword}

\end{frontmatter}

%% \linenumbers

%% main text
In the elastic coherent regime, thermoelectric transport through a nanosystem involves electric and heat currents between at least 
two electrodes, the source and the drain. One has a two-terminal setup where the electrons can be thermalized at different temperatures 
and chemical potentials in each electrodes. In the inelastic activated regime, a third terminal must be introduced, which contains the 
(quasi)-particles with which the electrons in the nanosystem interact. One often refers to this third terminal as the environment. 
For photovoltaic effects, it contains photons. For Mott variable range hopping (VRH) regime, it contains phonons. Such multi-terminal 
setups are characterized by two Fermi-Dirac distributions describing the thermalized source and drain, and a Bose-Einstein distribution 
describing the thermalized baths of photons or phonons. Markus B\"uttiker was known for having set the foundation of the theory of multi-terminal 
mesoscopic transport~\cite{Buttiker1} when he was at IBM Yorktown Heights. More recently, he studied multiterminal thermoelectric transport. 
For instance, he considered the case where a third electron reservoir is capacitively coupled with a nanosystem contacting a source and a drain. 
In Ref.~\cite{Sanchez}, it was shown that an electrical current through the nanosystem can cool the third terminal or that heating the third terminal 
can give rise to an electrical current through the nanosystem. This last effect was recently measured in Saclay~\cite{Glattli} and Wurzburg~\cite{Hartmann}. 
In Ref.~\cite{Sanchez}, the considered setup involves three Fermi-Dirac distributions and three currents: the electric and heat currents 
between the source and the drain, and the heat current between the electrons inside the nanosystem and those inside the third terminal. 
As emphasized~\cite{Jiang2012,Jiang2013} by Yoseph Imry, hopping transport is also a genuine three-terminal transport involving two electron 
baths and a phonon bath. This is why we feel appropriate to publish this study of multi-terminal activated transport in a volume in memory of 
Markus B\"uttiker. 

\begin{figure}
    \begin{center}
            \includegraphics[width=0.9\columnwidth]{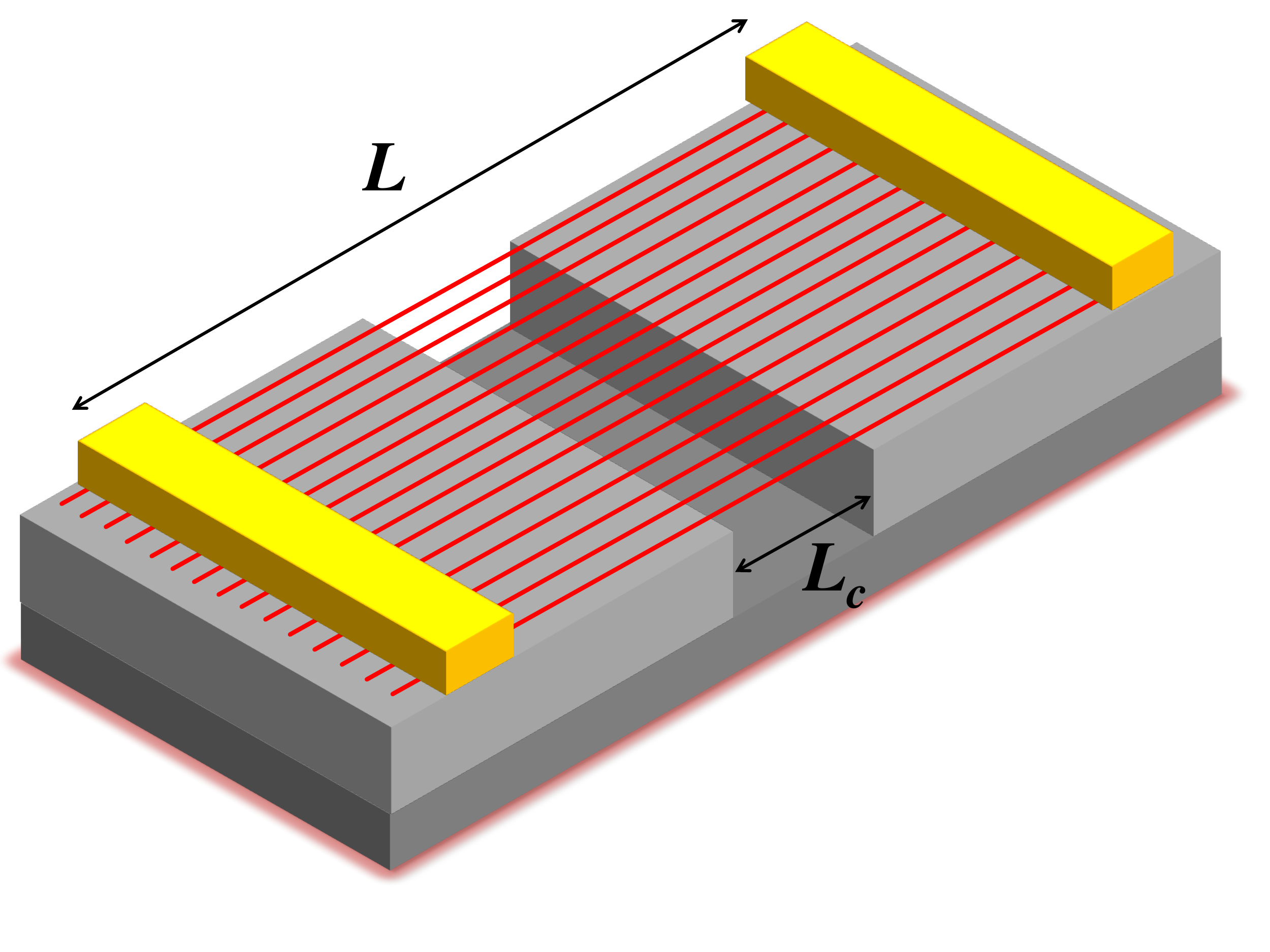}
    \end{center}
    \caption{%
Thin doped semiconductor NWs (red) are deposited in parallel on an insulating substrate (light grey). Two metallic electrodes 
(yellow) provide a source and a drain. An electron current flows through the NWs when a bias voltage is applied, while the 
NW conduction bands are shifted by a voltage applied on a back gate (darker grey) put below the substrate. 
The substrate is cut in two parts, the middle of the NWs of length $L$ being suspended over a small length $L_c$. This defines 
the source and drain sides of the substrate.
     }%
\label{FIG1a}
\end{figure}

\begin{figure}
    \begin{center}
            \includegraphics[width=0.9\columnwidth]{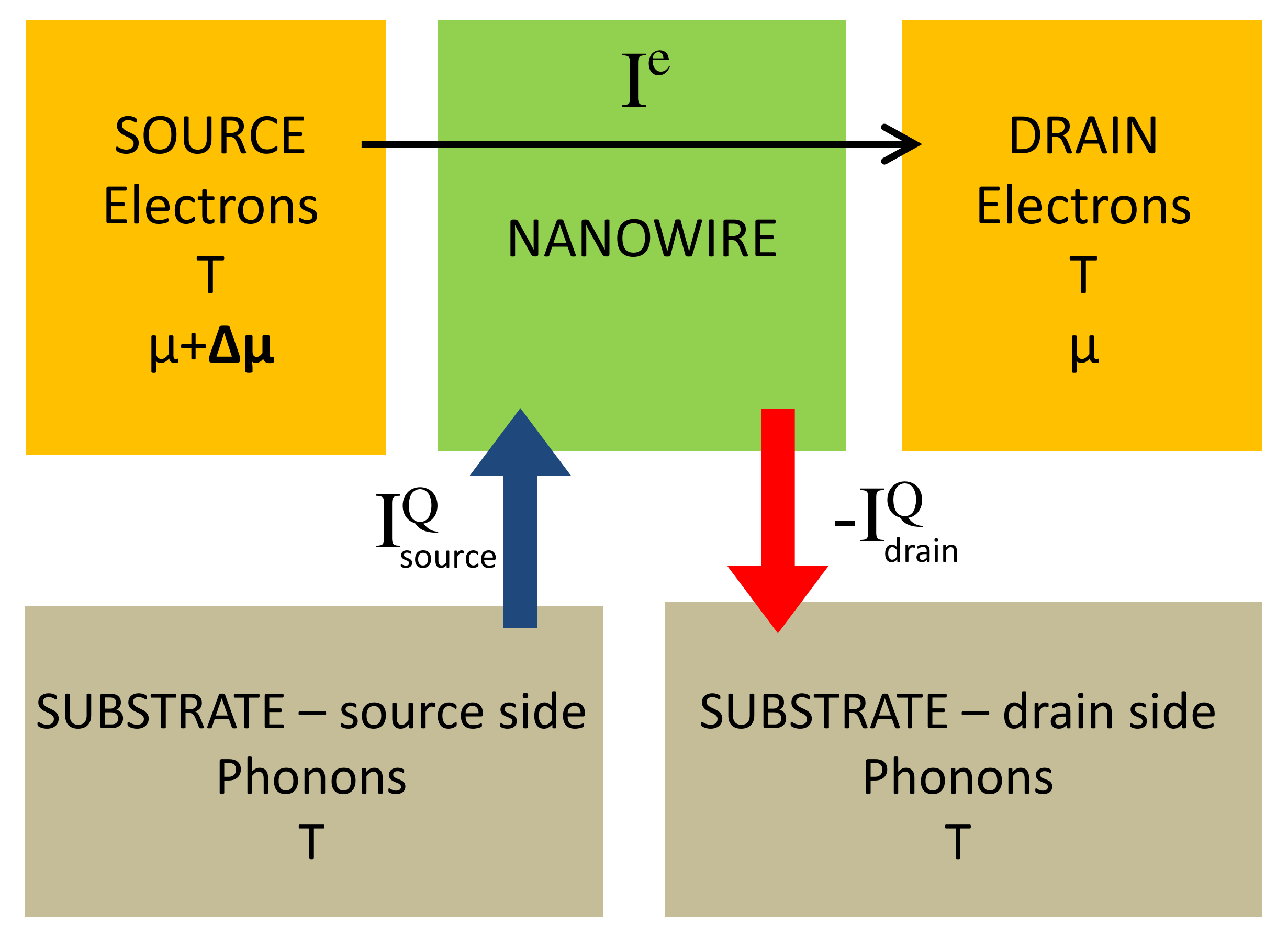}
    \end{center}
    \caption{%
Multi-terminal setup corresponding to phonon-assisted transport in a single nanowire. It involves two electron reservoirs 
(source and drain) and a phonon bath (provided for a 1d MOSFET by the insulating oxide substrate on which 
the NW is deposited). The substrate is divided by a small cut between its source and drain sides. 
 }%
\label{FIG1b}    
\end{figure}

 We consider the array of one dimensional (1d) metal-oxide-semiconductor-field-effect-transistors (MOSFETs) shown in Fig.~\ref{FIG1a}. 
The array is made of many thin doped semiconductor nanowires (NWs) arranged in parallel. Activated transport in each of these NWs 
corresponds to the multi-terminal setup sketched in Fig.~\ref{FIG1b}. The source and the drain are the two electron reservoirs, 
whereas the substrate, divided between its source and drain sides, plays the role of a phonon reservoir. The electrons are thermalized 
 in the source and the drain with Fermi-Dirac distribution of temperature $T_{S,D}$ and chemical potential $\mu_{S,D}$. The phonons are 
thermalized in the substrate with a Bose-Einstein distribution of temperature $T_{ph}$. 
If the electron states are localized inside the NWs, but also coupled with the phonons of the substrate, electron transport through the 
NWs becomes activated when their lengths exceed the Mott hopping length $L_{M}$ and is governed by Mott's variable range hopping (VRH) 
mechanism. Hereafter, we set $\mu_{D}=\mu= 0$ and $\mu_{S}=\delta \mu =e \delta V$ where $e$ is the electron charge and $\delta V$ 
the bias voltage. We assume a uniform temperature $T_{S}=T_{D}=T_{ph}$. The bias induces an electron current $I^{e}$ from the 
source to the drain. In an activated regime, $I^{e}$ will be also due to the presence of local heat currents $I^Q(x)$ between the 
nanowire and the substrate. If the nanowire density of states $\nu(E)$ and localization length $\xi(E)$ do not depend on the energy $E$ in 
the vicinity of $\mu$, $I^Q(x)$ fluctuates randomly as a function of the coordinate $x$ along the NW, a phonon of the substrate being sometimes 
absorbed, sometimes emitted. As a consequence, the ensemble average value of the heat exchange vanishes, $<I^Q(x)>=0$. Interesting effects occur instead 
if $\nu(E)$ and $\xi(E)$ are strongly energy dependent. Particle-hole symmetry is broken: This gives rise to large thermoelectric 
effects~\cite{Bosisio20142,Bosisio2015} and the local heat currents along the NWs become finite, $<I^Q(x)> \neq 0$, the phonons being mainly absorbed 
or emitted near the NW boundaries. 
\\
We study such a case using for the density of states per unit length $\nu(E)$ and for the localization length $\xi(E)$ analytical results 
describing a 1d lattice with random site potentials in the weak disorder limit. Shifting the NW conduction band with the gate voltage $V_g$ such 
that the electrons are injected from the source around its lower edge, the phonons are mainly absorbed near the source and emitted near the drain. 
This corresponds to the sketch given in Fig.~\ref{FIG1b}. Inducing with $\delta \mu$ an electric current $I^e$, one obtains a heat current 
$I^Q_{source}>0$ cooling the source side of the substrate, while $I^Q_{drain}<0$ heats the drain side. The cooling/heating effects are reversed if 
the electrons are injected around the upper band edge and vanish if the injection is made around the band center. Phonon absorption/emission 
can thus be controlled with bias and gate voltages in 1d MOSFETs.  

\section{VRH transport with energy independent localization length $\xi$ and density of states $\nu$}

Phonon assisted hops between localized states and the associated mechanism of VRH transport have been mainly studied when the energy dependence 
of the density of the localized states per unit length $\nu(E)$ and of their localization length $\xi(E)$ can be neglected around $\mu$. Under 
this assumption, the electron transfer from one localized state to another one separated by a distance $x$ and an energy $\delta E \propto 1/(\nu x)$ 
results from the competition between the elastic tunneling probability ($\propto \exp\{-2x/\xi\}$) to do a hop of length $x$ in space and the Boltzmann 
probability ($\propto \exp\{-\delta E/k_BT\}$) to do a hop of $\delta E$ in energy. This competition gives rise to an optimal hopping length $L_M$, 
the Mott's hopping length, and an optimal hopping energy $\Delta_M$. In one dimension, one has
\be
\label{L_Mott}
L_M= \sqrt{\frac{\xi}{2 \nu k_BT}},
\ee 
\be
\label{Delta_Mott}
\Delta_M= 2\sqrt{\frac{\xi k_BT}{2\nu}}.
\ee
Electronic transport is achieved via several hops of magnitude $\approx L_M$ (with $\xi<L_M<L$) in space or $\Delta_M$ in energy, 
and the conductance $G$ can be expressed in terms either of $L_M$ or $\Delta_M$:
\be
\label{conductance_VRH}
G \propto \text{exp}\left\{ -\frac{2L_M}{\xi}\right\}=\text{exp}\left\{-\frac{\Delta_M}{k_BT}\right\}.
\ee
Since $L_M$ is a decreasing function of the temperature, the VRH regime takes place above the activation temperature
\be
\label{T_Activation}
k_BT_x=\frac{\xi}{2\nu L^{2}},
\ee
at which $L_M\simeq L$ and below the Mott temperature
\be
\label{T_Mott}
k_BT_M=\frac{2}{\nu \xi}, 
\ee
at which $L_M\simeq \xi$. Below $T_x$, $L_M$ exceeds the system size and electron transport becomes elastic and coherent 
(see Ref.~\cite{Bosisio20141}). Above $T_M$, $L_M \leq \xi$ and transport is simply activated between nearest neighbor 
localized states. Actually, the crossover from VRH to simply activated transport takes place at temperatures lower than 
$T_M$ in one dimension. This is due to the presence of highly resistive regions in energy-position space, where 
electrons cannot find available states at distances $\sim \Delta_M, L_M$. These regions behave as "breaks" that electrons are 
constraint to cross by thermal activation, resulting in a simply activated temperature dependence of the overall 
resistance~\cite{Kurkijarvi1973,Raikh1989}.  

  A more microscopic approach for deriving the above expressions consists in replacing the transport problem by a random-resistor 
network one~\cite{Miller1960} in which the hopping between two localized states $i$ and $j$ is effectively described by a resistor 
$\rho_{ij}$. As proposed by Ambegaokar, Halperin and Langer, one can approximate~\cite{Ambegaokar1971}  the resistance of the network 
by the lowest resistance $\rho_c$ such that the resistors $\rho_{ij}<\rho_c$ form a percolating network traversing the sample. In one 
dimension, an approximation for $\rho_c$ was derived in Ref.~\cite{Serota1986}
\begin{equation}
\ln\rho_c = \left(\frac{T_M}{2T}\right)^{1/2} \left\{ \ln\left[ \frac{2L}{\xi} 
\left(\frac{2T}{T_M}\right)^{1/2} \left(\ln\frac{2L}{\xi}\right)^{1/2}\right] \right\}^{1/2},
\label{eq_rhoc}
 \end{equation}
which predicts a $T^{-1/2}$ behavior independently of the length $L$ when $T \to 0$, and a simple activated ($T^{-1}$) behavior for 
wires longer than $\xi/2 \sqrt{T_M/T} \exp\{T_M/T\}$.

\section{VRH transport in realistic thin nanowires}

\subsection{Energy dependence of the localization length $\xi(E)$ and density of states $\nu(E)$}

All the previous results were based on the assumption that $\xi(E)$ and $\nu(E)$ are independent of the energy. 
Such an assumption becomes questionable in one dimension, notably as one approaches the band edges. Let us take 
the widely used Anderson model: A one dimensional lattice of $N=L/a$ sites with lattice spacing $a=1$, nearest 
neighbor hopping $t$, and random site potentials $\epsilon_i$ uniformly distributed in the interval $[-W/2,W/2]$. 
Its energy levels $E_i$ are distributed in the energy interval $[E_c^{-},E_c^{+}]$, where $E_c^{\pm}=\pm(2t+W/2)$ 
defines the edges of the NW conduction band when $L\to\infty$. In the bulk of the band (i.e. for energies 
$|E|\lesssim 1.5t$), the density $\nu(E)$ of the Anderson model for small values of $W$ and $L \to \infty$ can be 
described by the formula (valid without disorder)
\be
\label{eq_dstOfStateBulk}
\nu_b(E)=\frac{1}{2\pi t\sqrt{1-(E/2t)^2}}~.
\ee
As one approaches the edges $E_c^{\pm}$, the disorder effects cannot be neglected and $\nu(E)$ is given by the analytical 
formula obtained~\cite{Derrida1984} by Derrida and Gardner around $E_c^{\pm}$:
\be
\label{eq_dstOfStateEdge}
\nu_e(E)=\sqrt{\frac{2}{\pi}}\left(\frac{12}{tW^2}\right)^{1/3}\frac{\mathcal{I}_1(X)}{[\mathcal{I}_{-1}(X)]^2},
\ee
where $X=(|E|-2t)t^{1/3}(12/W^2)^{2/3}$ and 
\be
\label{eq_integralIn}
\mathcal{I}_n(X)=\int_0^{\infty} y^{n/2}\,e^{-\frac{1}{6}y^3+2Xy}\,dy\,.
\ee
Similarly, the localization length $\xi(E)$ of the Anderson model can be described by analytical expressions 
valid for weak disorder. Inside the bulk of the band, one gets 
\be
\label{eq_xsi_bulk}
\xi_b(E)=\frac{24}{W^2}\left(4t^2-E^2\right)\,,
\ee
whereas 
\be
\label{eq_xsi_edge}
\xi_e(E)=2\left(\frac{12t^2}{W^2}\right)^{1/3}\frac{\mathcal{I}_{-1}(X)}{\mathcal{I}_{1}(X)}
\ee
as $E$ approaches the band edges $\pm 2t$. Figures of the functions $\nu(E)$ and $\xi(E)$ can be found in 
Ref.~\cite{Bosisio20141}, where the weak disorder expansions are shown to fit the numerically calculated 
values of $\nu(E)$ and $\xi(E)$ for values of the disorder parameter $W$ as large as $t$. 

\subsection{Energy dependence of the VRH scales}

\begin{figure}
    \begin{center}
            \includegraphics[width=0.9\columnwidth]{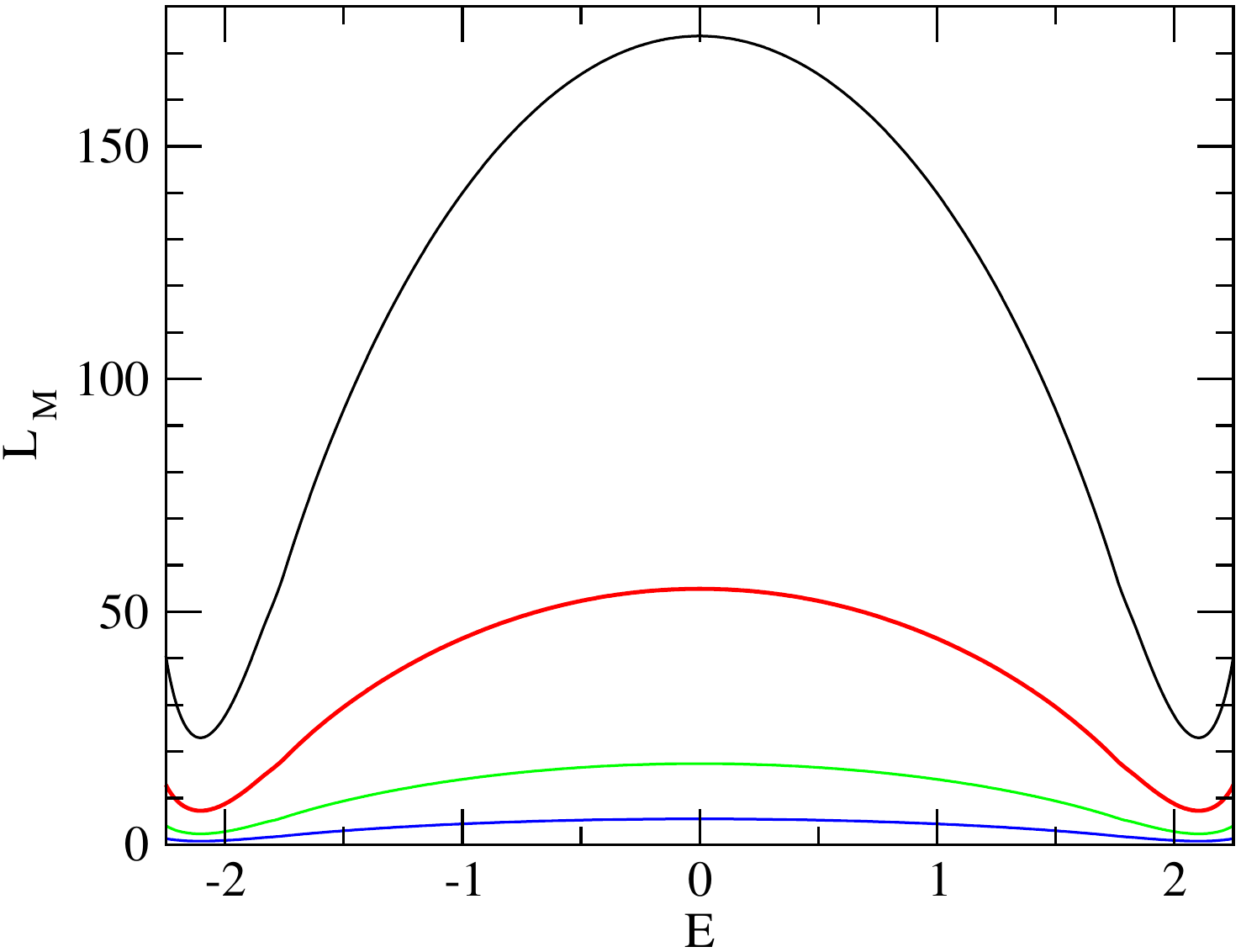}
    \end{center}
    \caption{Mott hopping length $L_M=\sqrt{\xi/(2\nu T)}$ as a function of $E$ for different temperatures (from top to bottom, 
$k_BT=0.01t$, $0.1t$, $t$ and $10t$). $\nu(E)$ and $\xi(E)$ are taken from the 1d Anderson model. NW length $L=1500$ and $W=t$.
}
\label{LM}
\end{figure}

\begin{figure}
    \begin{center}
            \includegraphics[width=0.9\columnwidth]{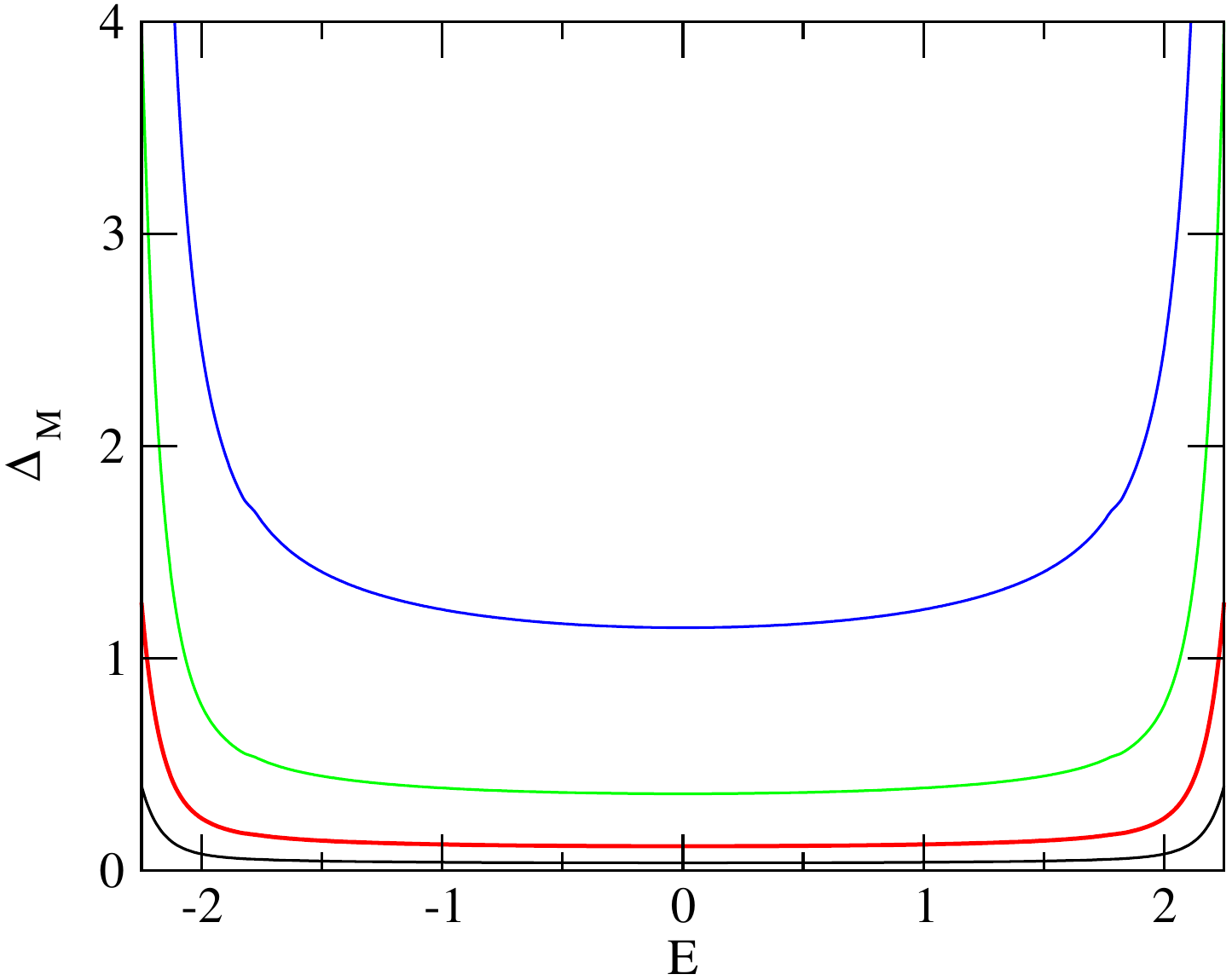}
    \end{center} 
    \caption{Mott hopping energy $\Delta_M=k_B\sqrt{T T_M}$ as a function of $E$. Same parameters and color code as in Fig.~\ref{LM}.
		}
		\label{DELTAM}
\end{figure}

\begin{figure}
    \begin{center}
            \includegraphics[width=0.9\columnwidth]{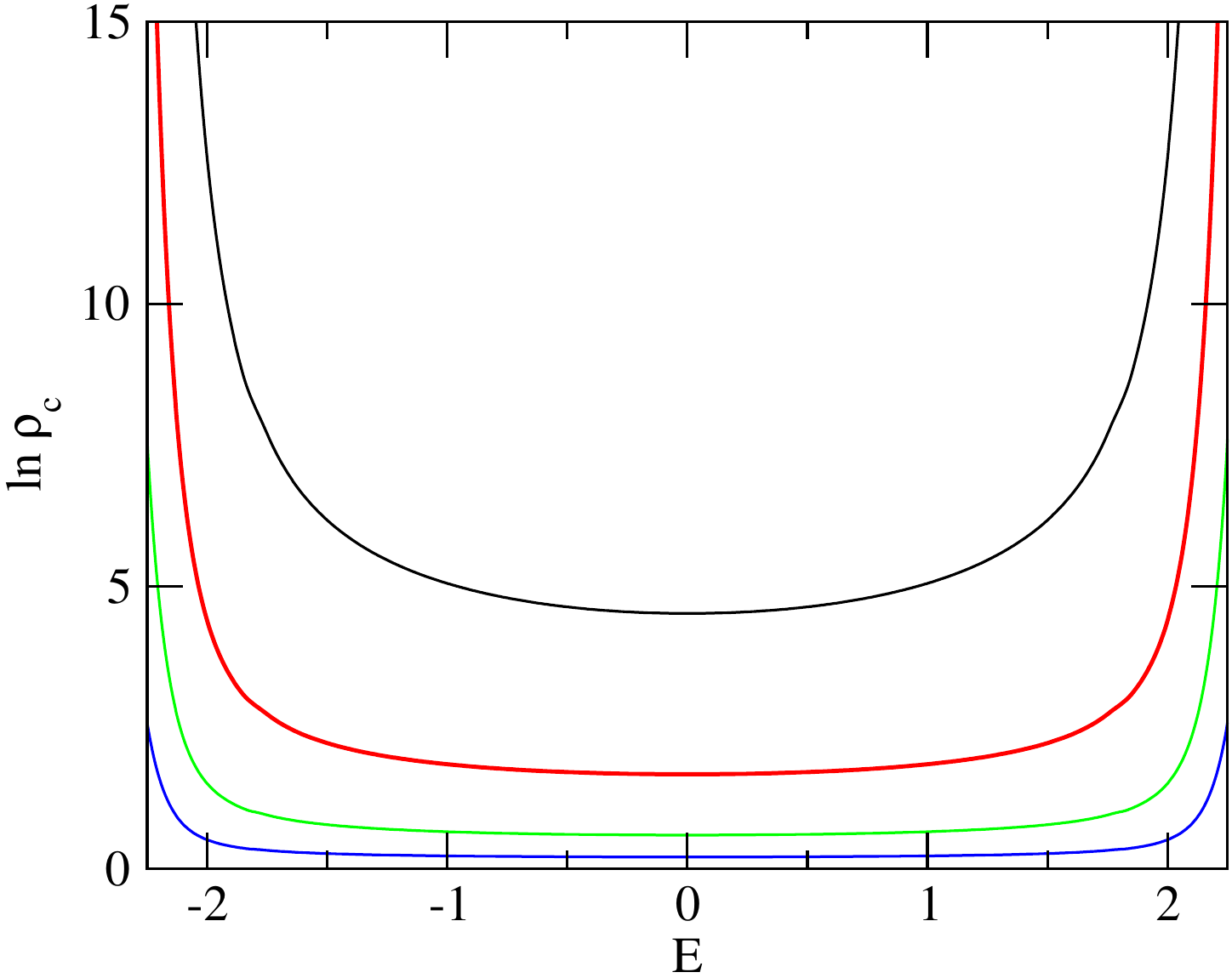}
    \end{center}
    \caption{Logarithm of the resistance $\ln\rho_c$ given by Eq.~\eqref{eq_rhoc} as a function of $E$. Same parameters and color code as in Fig.~\ref{LM}.
		}
		\label{rhoc}
\end{figure}

 Using the functions $\nu(E)$ and $\xi(E)$ of the 1d Anderson model and Eq.~\eqref{L_Mott}, Eq.~\eqref{Delta_Mott} and Eq.~\eqref{eq_rhoc}, 
we show how the Mott hopping length $L_M(E)$ (Fig.~\ref{LM}) and the hopping energy $\Delta_M(E)$ (Fig.~\ref{DELTAM}) vary with $E$. 
The energy dependence of the resistance $\rho_c$ which characterizes the best percolation network describing the random resistor network 
is shown in Fig.~\ref{rhoc}. 

Strictly speaking, these functions $L_M(E)$, $\Delta_M(E)$ and $\rho_c(E)$ do not characterize hopping transport in a \emph{single} Anderson model, 
but in an ensemble of 1d models where $\nu$ and $\xi$ do not depend on $E$, but rather take uniform values which are given by those of the  
Anderson model at an energy $E$. Nevertheless, one can easily guess from the figures of $L_M(E)$, $\Delta_M(E)$  and $\rho_c(E)$ 
how an electron will cross the nanowire. If it is injected from the source electrode at an energy around the 
 center of the conduction band, it should exchange much less energy with the substrate (see Fig.~\ref{DELTAM}) and will   
find much smaller resistances $\rho_c$ (see Fig.~\ref{rhoc}) than if it is injected at an energy near the NW band edges. 
Let us take an electron injected near the lower band edge. It is likely than the percolation path that it will follow  
will consist first in absorbing many phonons on the NW side close to the source, to reach higher energies where the resistance 
$\rho_c$ is weaker (see Fig.~\ref{rhoc}). Then, it will continue around an optimum energy where it will jump from one localized state 
to another by emitting or absorbing phonons at random. When it will reach the NW drain side, it will mostly emit phonons, in order 
to reach an energy of the order of the chemical potential of the drain electrode.  

\section{Numerical study of the random resistor network}

  To proceed further, we numerically solve the Miller-Abrahams resistor network~\cite{Miller1960} which 
describes hopping transport. The nodes of the network are given by the NW localized states. Each pair 
of nodes $i, j$ is connected by an effective resistor, which depends on the transition rates $\Gamma_{ij}, \Gamma_{ji}$ 
induced by local electron-phonon interactions. For a pair of localized states $i$ and $j$ of energies $E_i$ and $E_j$, 
Fermi golden rule~\cite{Jiang2013} gives:
\be
\label{eq:gamma_ij}
\Gamma_{ij}=\gamma_{ij}\,f_i\,(1-f_j)\,\left[N_{ij}+\theta(E_i-E_j)\right]\,,
\ee
where $f_i$ is the average occupation number of state $i$ and $N_{ij}=[\text{exp}\{|E_j-E_i|/k_BT\}-1]^{-1}$ 
is the phonon Bose distribution at energy $|E_j-E_i|$.  The presence of the Heaviside function accounts for the 
difference between phonon absorption and emission~\cite{Ambegaokar1971}. $\gamma_{ij}$ is the hopping probability 
$i \to j$ due to the absorption/emission of one phonon when $i$ is occupied and $j$ is empty. Neglecting the 
energy dependence of $\xi$ and $\nu$, one obtains in the limit $x_{ij}\gg\xi$
\be
\label{eq:gamma_ij_2}
\gamma_{ij}\simeq \gamma_{ep}\,\text{exp}(-2x_{ij}/\xi)\,.
\ee
Here $x_{ij}=|x_i-x_j|$ is the distance between the states, whereas  $\gamma_{ep}$, containing the electron-phonon 
interaction matrix element, depends on the electron-phonon coupling strength and the phonon density of states. Since it is weakly 
dependent on $E_i$, $E_j$ and $x_{ij}$ compared to the exponential factors, it is assumed to be constant. 
From hereon we will take $\gamma_{ep}=t/\hbar$ independent of the position along the nanowire
\footnote{This is justified if the length of the cut $L_c$ is very small (see Fig. \ref{FIG1a}): This limit will be assumed to hold 
in all the numerical simulations that follow.}.
Under the widely used approximation~\cite{Ambegaokar1971,Pollack1972,Shklovskii1984,Zvyagin1973} $|E_{ij}| \gg k_BT$, Eq.~\eqref{eq:gamma_ij} 
reduces to:
\be 
\label{eq:gamma_ij_3}
\Gamma_{ij}\simeq \gamma_{ep}\,e^{-2x_{ij}/\xi}\,e^{-(|E_i-\mu|+|E_j-\mu|+|E_i-E_j|)/2k_BT}\,.
\ee
However, we have seen that in one dimension, the energy dependence of $\xi$ and $\nu$ cannot be neglected. Hereafter, instead of 
using the approximation~\eqref{eq:gamma_ij_3}, we will use the exact expression~\eqref{eq:gamma_ij} and we will take for $\gamma_{ij}$ 
\be
\gamma_{ij}= {\gamma_{ep}} A(\xi_i,\xi_j,r_{ij}),
\label{eq:gamma_ij_app_4}
\ee
instead of Eq.~\eqref{eq:gamma_ij_2}, where 
\begin{align*}
A(\xi_i,\xi_j,r_{ij}) &= \left( 1/\xi_i - 1/\xi_j\right)^{-2} \left[\frac{\exp\{-2r_{ij}/\xi_j\}}{\xi_i^2}\right. \\ 
& + \left.\frac{\exp\{-2r_{ij}/\xi_i\}}{\xi_j^2} -\frac{2\exp\{-r_{ij} (1/\xi_i + 1/\xi_j)}{\xi_i \xi_j}\right].  
\end{align*}
 Eq.~\eqref{eq:gamma_ij_app_4} takes into account the energy dependence of $\nu(E)$ and $\xi(E)$ and 
is derived in Ref.~\cite{Bosisio20142}. It reduces to Eq.~\eqref{eq:gamma_ij_2} when $\xi_i \to \xi_j=\xi$ and 
(to leading order) when $\xi_i=\xi \gg \xi_j$.
\\ 
\indent The transition rates between each state $i$ and the contacts $\alpha$ ($\alpha=L$ for the source or $R$ 
for the drain) are assumed do be dominated by elastic tunneling contributions (see Refs.~\cite{Jiang2012,Jiang2013}) and read: 
\be
\label{eq:gamma_il}
\Gamma_{i\alpha}=\gamma_{i\alpha}\,f_i\,\left[1-f_{\alpha}(E_i)\right],
\ee
where
\be
\gamma_{i\alpha}\simeq\gamma_{e}\,\text{exp}(-2x_{i\alpha}/\xi_i)\,.
\ee
In the above equations $f_{\alpha}(E)=[\text{exp}\{(E-\mu_\alpha)/k_BT\}+1]^{-1}$ is the contact $\alpha$'s
Fermi-Dirac distribution, $x_{i\alpha}$ denotes the distance of the state $i$ from $\alpha$, 
and $\gamma_{e}$ is a rate quantifying the coupling between the localized states and the contacts. In the following numerical simulations, we will assume $\gamma_e=t/\hbar$.\\
\indent Then, the net electric currents flowing between each pair of localized states and 
between states and contacts read
\begin{subequations}
\label{eq:currents}
\begin{align}
I_{ij} &= e\,(\Gamma_{ij}-\Gamma_{ji})\label{eq:currents_1},\\
I_{i\alpha} &= e\,(\Gamma_{i\alpha}-\Gamma_{\alpha i}),\,\qquad\,\alpha=L,R\label{eq:currents_2}
\end{align}
\end{subequations}
$e<0$ being the electron charge. Imposing current conservation through all the $L$ nodes $i$ of the network 
$ \sum_{j} I_{ij}+\sum_{\alpha}I_{i\alpha}=0$, one obtains $L$ coupled linear equations. Solving numerically 
this set of equations in the limit where the bias voltage $\delta \mu \to 0$ and where the temperature 
is $T$ everywhere gives the unknown occupation numbers $f_i$ of the localized states $i$ and hence the linear 
response solution of the random resistor network problem (see for more details Refs.~\cite{Jiang2013,Bosisio20142}). 
This allows us to obtain all the electrical currents $I_{ij}$ and $I_{i\alpha}$.

The set of energies $E_i$, of positions $x_i$ and of localization lengths $\xi_i$ are required as input parameters 
of the random resistor network problem. Hereafter, we use a simplified model as it is conventional in numerical 
simulations of VRH transport (see~\cite{Lee1984,Serota1986,Jiang2013}): The $E_i$ are uncorrelated variables 
taken with a distribution corresponding to the density $\nu(E)$ of the 1d Anderson model [Eqs.~\eqref{eq_dstOfStateBulk} 
and~\eqref{eq_dstOfStateEdge}], while the corresponding localization lengths $\xi(E_i)$ are taken using 
Eqs.~\eqref{eq_xsi_bulk} and~\eqref{eq_xsi_edge}. The positions $x_i$ are taken at random (with a uniform distribution) 
between $0$ and $L$ along a chain of length $L=Na$, $N$ being the number of sites with spacing $a=1$.

\section{Heat exchange between substrate phonons and NW electrons}

\subsection{Local heat currents}  
\begin{figure}
    \begin{center}
            \includegraphics[width=\columnwidth]{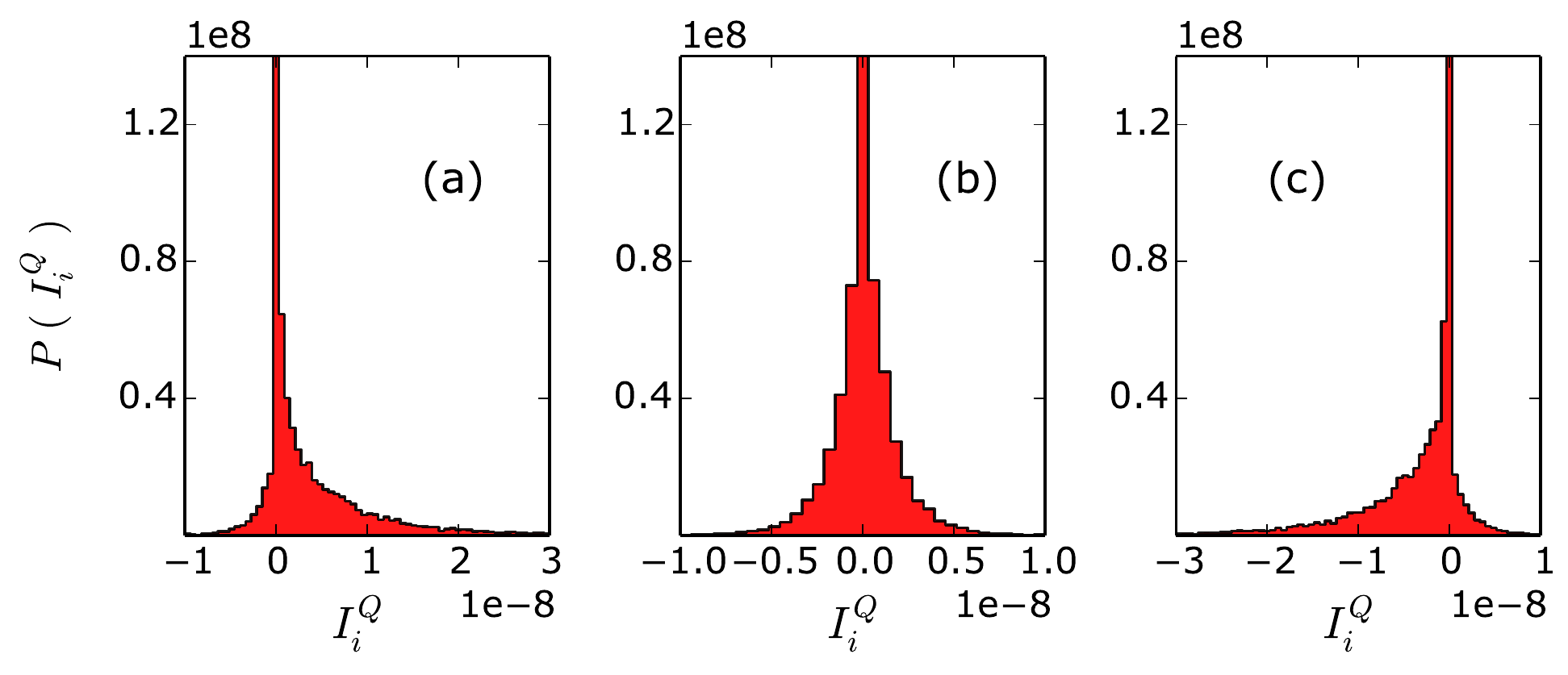}
    \end{center}
    \caption{Probability distribution of $I^Q_i=\sum_j I^Q_{ij}$ for 3 positions $x_0$ along a nanowire [$x_0=10$ (left), $750$ (middle) and 
$1491$ (right)] of length $L=1500$. Statistics over $2000$ NWs, obtained by counting the number of "points" in $[x_0-dx/2,x_0+dx/2]$, 
with $dx=20$. Parameters: $V_g=2.25t$, $k_BT=0.1t$, Anderson model with $W=t$, $\delta\mu=10^{-5}t$.}
    \label{fig_1DhistoIQi_T0.5}
\end{figure}

 Let us consider a pair of localized states $i$ and $j$, of respective energies $E_i$ and $E_j$. The heat current absorbed from 
(or released to) the substrate phonon bath by an electron in the transition $i\to j$ reads $I_{ij}^{Q}=\left(E_j-E_i\right)I^N_{ij}$, 
$I^N_{ij}=\Gamma_{ij}-\Gamma_{ji}$ being the hopping particle current between $i$ and $j$.\cite{Bosisio20142} The local heat current 
associated to a given localized state $i$ is given by summing over all the possible hops with the other NW states $j$:
\be\label{eq:IQ_i}
I_{i}^{Q}=\sum_j I_{ij}^Q = \sum_{j} \left(E_j-E_i\right)I^N_{ij}.
\ee
We take the convention that $I^Q_{i}$ is positive (negative) when it enters (leaves) the NW at site $i$.  The probability 
distribution of $I_{i}^{Q}$ is shown in Fig.~\ref{fig_1DhistoIQi_T0.5} for states $i$ located near three different positions $x_0$ 
along the NW. The gate potential ($V_g=2.25t$) has been chosen such that the chemical potential $\mu$ probes the lower band edge of the 
NW conduction band. Near the source, the distribution is highly asymmetric: The electrons injected in an energy interval of width 
$k_BT=0.1t$ around the lower band edge need to absorb phonons for reaching higher energies $E$ where the paths of the random resistor 
network are less resistive. Far from the NW boundaries, the distribution becomes symmetric: The electrons have found an optimum value 
of $E$ around which they stay, emitting or absorbing phonons at random. Near the drain, the distribution becomes again asymmetric, 
as the electrons emit more phonons to decrease their energies to reach the chemical potential $\mu$ of the drain.    

 \begin{figure}
    \begin{center}
            \includegraphics[width=\columnwidth]{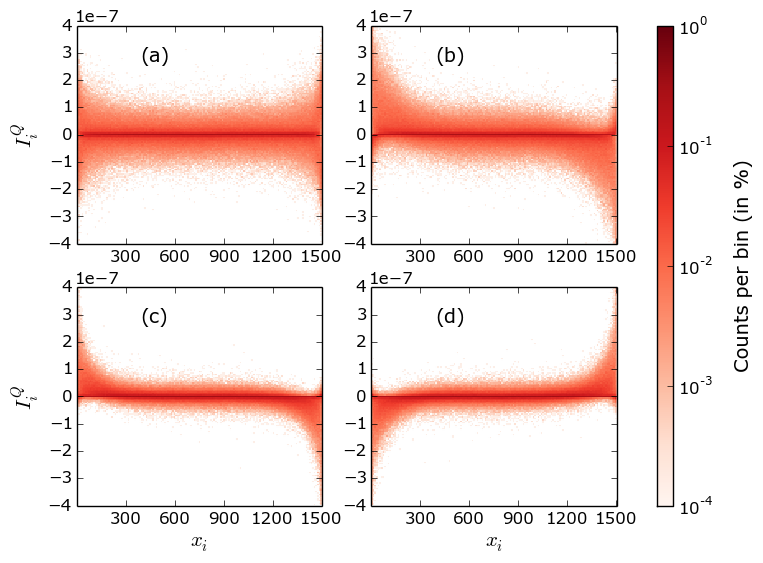}
    \end{center}
    \caption{%
2d histograms giving the distribution of local heat currents $I^Q_i=\sum_j I^Q_{ij}$ as a function of the position $x_i$ along the NW, 
calculated for a gate potential of value $V_g=0$ (a), $V_g=t$ (b), $V_g=2.25t$ (c), and $V_g=-2.25t$ (d). 
Parameters: $L=1500$, $W=t$, $k_BT=0.5t$, $\delta\mu=10^{-5}t$, statistics over $500$ NWs.    		
        }%
\label{2Dhistograms_IQi_T0p5}
\end{figure}

In Fig.~\ref{2Dhistograms_IQi_T0p5} we show 2d histograms of the local heat currents $I^Q_i$ as a function of the position $x_i$ inside 
the NW. The data have been calculated for a temperature $k_BT=0.5t$ and four different values of $V_g$, corresponding to electron injection 
at the band center ($V_g=0$), below the band center ($V_g=t$) and around the lower ($V_g=2.25t$) and upper ($V_g=-2.25t$) band edges of the NW 
conduction band, respectively.
At the band center, the fluctuations of the local heat currents are symmetric around a zero average. They are larger near the NW 
boundaries and remain independent of the coordinate $x_i$ otherwise. Away from the band center, one can see that the fluctuations are not symmetric 
near the NW boundaries, though they become symmetric again far from the boundaries. When the electrons are injected through the NW in the lower 
energy part of the NW band, more phonons of the substrate are absorbed than emitted near the source electrode. The effect is reversed when the 
electrons are injected in the higher energy part of the band (by taking a negative gate potential $V_g$): It is now near the drain that the phonons 
are mainly absorbed. The 2d histograms corresponding to $V_g=2.25t$ and $-2.25t$ are symmetric by inversion with respect to $I^Q_i=0$.

\begin{figure}
    \begin{center}
            \includegraphics[width=\columnwidth]{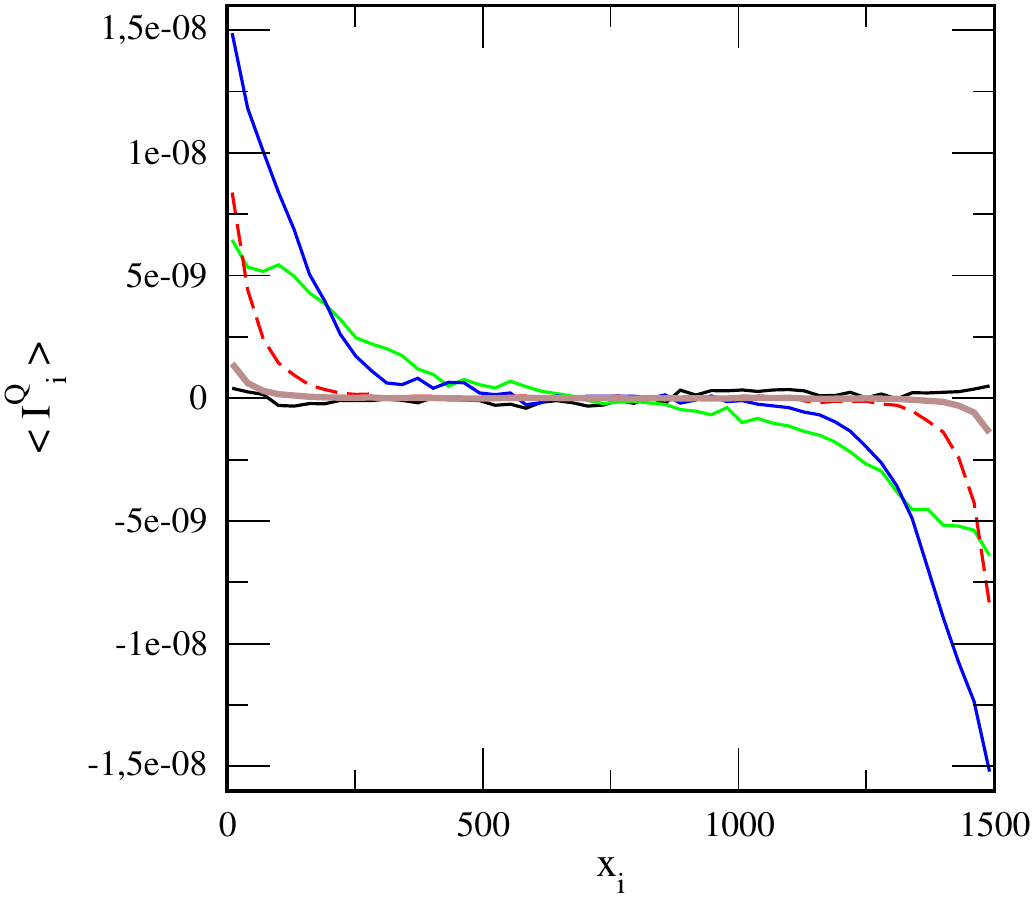}
    \end{center}
    \caption{%
    		Mean value $<I^Q_i>$ of the local heat currents $I^Q_i$ as a function of the position $x_i$ along the nanowire, for different values of 
$V_g$ ($V_g=0$ (black line), $t$ (green line), $1.5t$ (blue line), $2t$ (red dashed line), and $2.25t$ (thick brown line)). Parameters: 
$L=1500$, $W=t$, $k_BT=0.1t$, $\delta\mu=10^{-5}t$, statistics over $2000$ wires. Note that for each position $x_i$, we count the number of 
"points" in $[x_i-dx/2,x_i+dx/2]$, with $dx=20$ here.
        }%
    \label{fig_means_p1}
\end{figure}

In Fig.~\ref{fig_means_p1}, one can see how varies the mean value of the local heat current $I^Q_i$ along the nanowire for different values of $V_g$. 
Though the heat current extracted from the substrate near the source takes larger local values when one probes the band edges, one can see that 
it is more advantageous for cooling the source side of the substrate to inject the electrons at an intermediate energy, when $V_g\approx 1.5t$ 
(see also Sec.~\ref{sec:optimum}). 

\begin{figure}
    \begin{center}
            \includegraphics[width=\columnwidth]{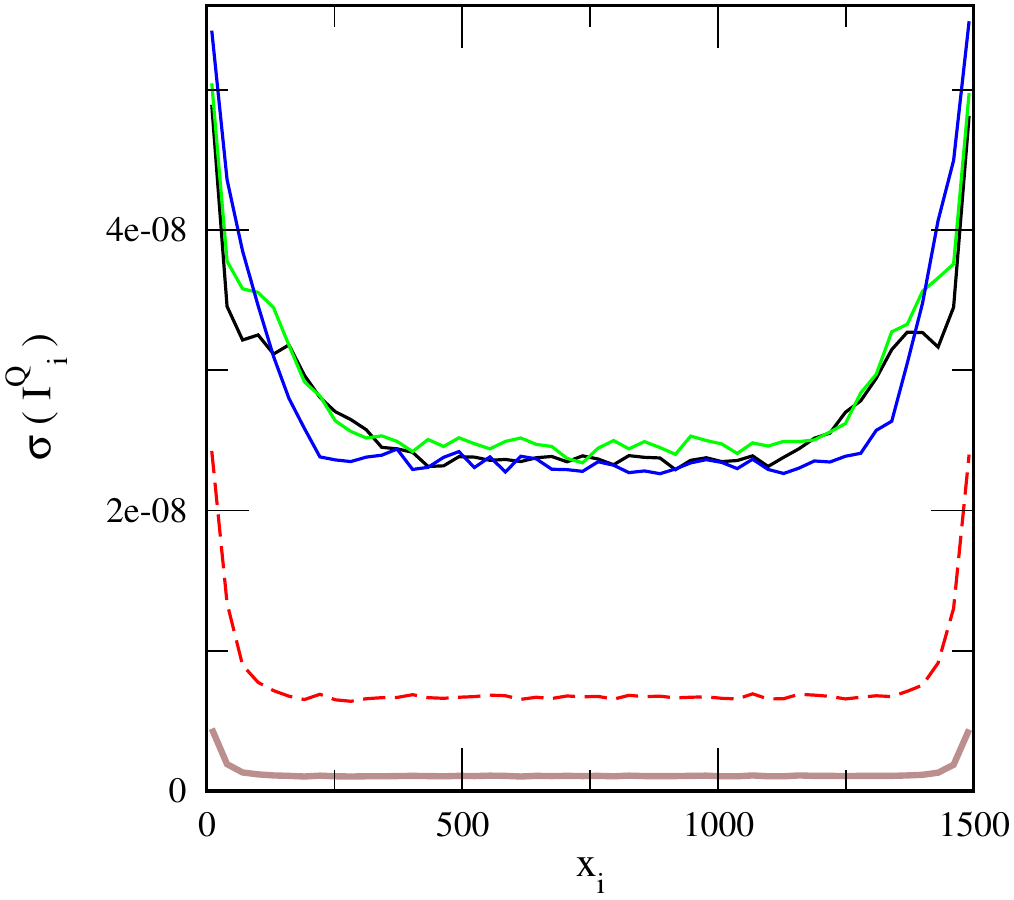}
    \end{center}
    \caption{%
    Standard deviation of the distribution of $I^Q_i$ as a function of the position $x_i$, obtained using the same parameters than in 
Fig.~\ref{fig_means_p1}.
        }%
\label{rms}
\end{figure}

In Fig.~\ref{rms}, one can see how the standard deviation of the local heat currents $I^Q_i$ varies as a function of the position 
$x_i$ along the nanowire, using the same parameters and color code than in Fig.~\ref{fig_means_p1}. The fluctuations of the local heat 
currents are strongly reduced when the electrons are injected near the band edges, and become independent of the injection energy 
if this energy is taken around the band center.

\subsection{Total heat current extracted from the source side of the substrate}

\begin{figure}
    \begin{center}
            \includegraphics[width=\columnwidth]{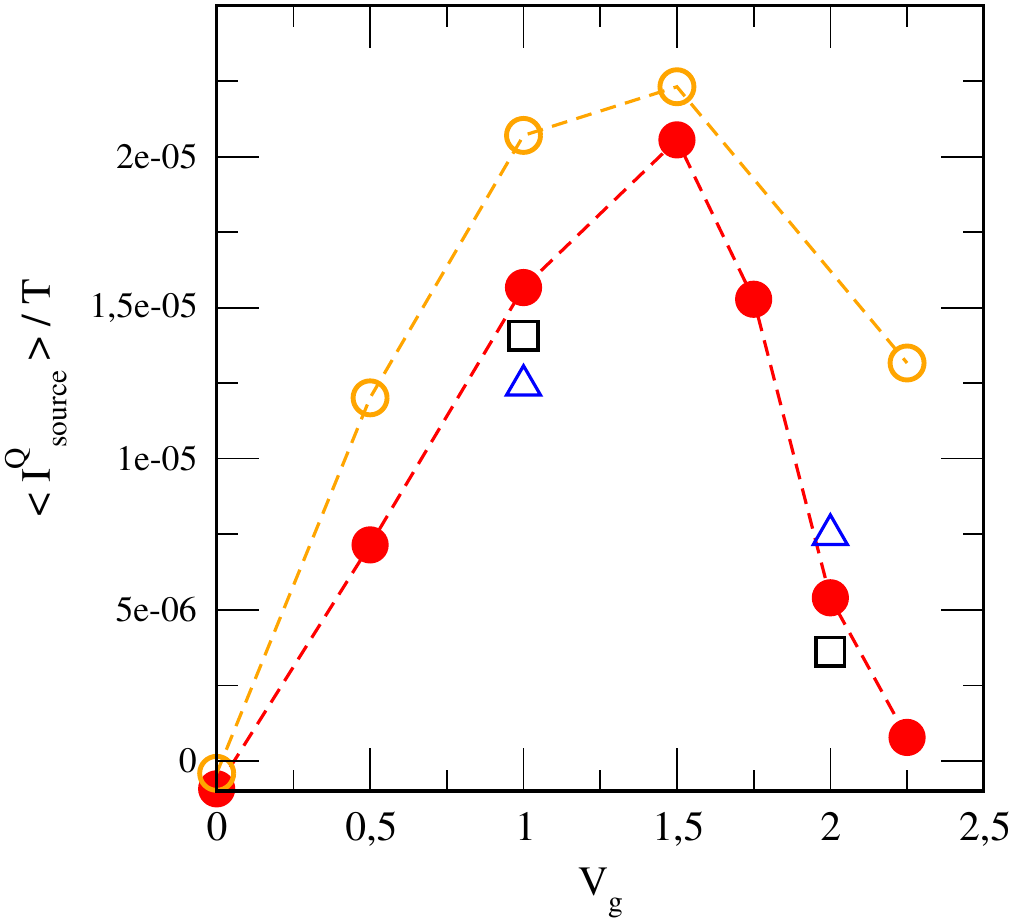}
    \end{center}
    \caption{%
   Total heat current $I^Q_{source}$ extracted from the source side of the substrate (divided by $k_BT$) as a function of the gate voltage $V_g$. 
Symbols correspond to: $L=1500$ and $k_BT=0.1t$ (full circles), $L=1500$ and $k_BT=0.5t$ (empty circles), $L=750$ and $k_BT=0.1t$ 
(triangles), $L=3000$ and $k_BT=0.1t$ (squares) Parameters: $W=t$, $\delta\mu=10^{-5}t$, statistics over $2000$ wires.
        }%
\label{IQsource}
\end{figure}

\begin{figure}
    \begin{center}
            \includegraphics[width=\columnwidth]{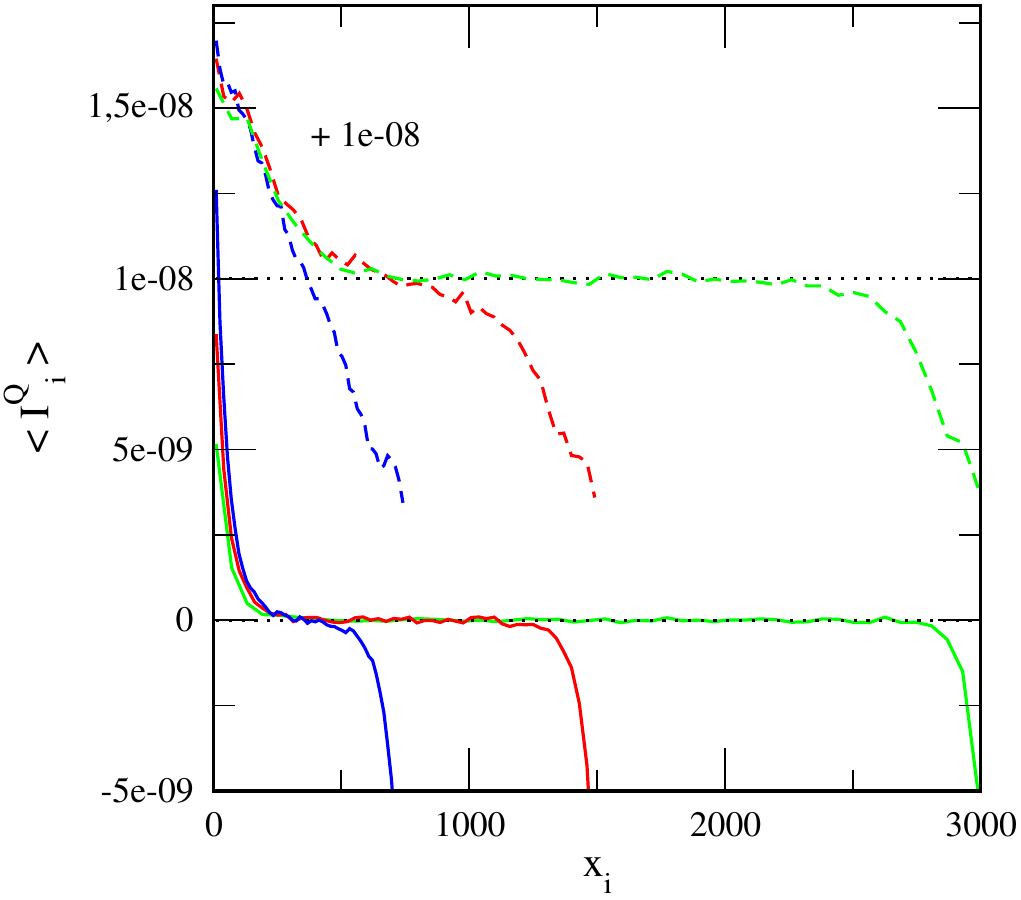}
    \end{center}
    \caption{%
    		Mean value $<I^Q_i>$ of the local heat currents $I^Q_i$ as a function of the position $x_i$ along the nanowire, 
for different lengths of the nanowire ($L=750$ (blue curves), $L=1500$ (red curves), and $L=3000$ (green curves)), 
and for two values of $V_g$ [$V_g=t$ (dashed lines) and $V_g=2t$ (full lines)]. Parameters: $W=t$, $k_BT=0.1t$, $\delta\mu=10^{-5}t$, 
statistics over $2000$ wires. Data for $V_g=t$ have been shifted upwards for a better visibility.
        }%
\label{asymtotics}
\end{figure}

As sketched in Fig.~\ref{FIG1a}, we assume that the substrate has a small cut over which the NW is suspended, 
and we are interested in cooling the substrate on one side of the cut, while the other side would be heated. 
Assuming that the contribution of the NW center can be neglected (or that the cut is very short), the total heat 
current extracted from the source side of the substrate using a NW of length $L$ is defined as 
\be
I^Q_{source}=\int_0^{L/2} <I^Q_x> dx,
\label{Isource}
\ee
i.e. $I^Q_{source}$ gives the area under the curves shown in Fig.~\ref{fig_means_p1}, taken from $0$ to $L/2$. $I^Q_{source}/k_BT$ 
is shown as a function of $V_g$ for $k_BT=0.1t$ and $0.5t$ in Fig.~\ref{IQsource}. One can see that the cooling effect is maximum 
when $V_g\approx 1.5t$. Very approximately, $I^Q_{source} \propto k_BT$ in the studied temperature domain and does not depend on $L$. 
The existence of an asymptotic regime when $L \to \infty$ is confirmed in Fig.~\ref{asymtotics}. The average local heat currents do 
not depend on $L$ near the NW boundaries and vanish in its middle.  

 \subsection{Optimum condition for phonon absorption by the first half of a NW}
\label{sec:optimum}

 In Fig.~\ref{IQsource}, one can see that $V_g\approx 1.5t$ gives the largest value for $I^Q_{source}$. This optimal value comes from 
the competition between two effects. On the one side, injecting electrons of energies $E$ near the lower band edge favors phonon absorption.  
On the other side, at these energies the resistance $\rho_c$ of the random resistor network is very large (see Fig.~\ref{rhoc}), at least 
near the NW boundaries. This necessarily reduces the particle current crossing the NW, and hence the heat current extracted from the source 
side of the substrate. For a few disordered chains, we have studied the map of currents larger than some threshold value which are driven 
by the bias voltage. Around the lower band edge ($V_g=2.25t$), it does not seem that the electrons are able to hop high enough in energy 
for reaching the band center where $\rho_c$ is minimum. Despite very large sample to sample fluctuations, the currents $I_{ij}^N$ connecting 
states of positions $i$ and $j$ located around the middle of the NW have energies $E_i$ and $E_j$ mainly located at a distance 
$\Delta E \approx 1.5t$ below the band center. $V_g \approx 1.5t$ could correspond to the largest energy distance over which the electrons 
can hop for reaching the band center in the middle of the NW. The difficulty of reaching energies far from the chemical potential $\mu$
can be understood by looking at the expression \eqref{eq:gamma_ij_3} for the transition rates. 
  
\subsection{Order of magnitude of heat fluxes using arrays of parallel NWs}  

 In Fig.~\ref{IQsource}, the total heat current integrated over the first half of a single NW is given in units of $t^2/\hbar$. 
We take the parameters values used in Ref.~\cite{Bosisio2015}: $t/k_B \approx 150$K gives $t \approx 12.9$meV.
This implies for $I^Q_{source}$ a maximum value around $0.4$pW when $V_g=1.5t$, $T=75$K and $\delta \mu = 10^{-5}t$. 
The used bias $\delta \mu/e$ corresponds to a very small voltage $\approx 0.13 \mu$V. Assuming that the non linear effects remain 
negligible by increasing the bias to a larger value, say $\delta V = 1.3$mV corresponding to $\delta \mu =0.1t$, one would 
have a maximum value $\approx 4$nW for a single NW of a length $\approx 2.2 \mu$m. This length corresponds to a lattice of 700 
sites with a spacing of $3.2$nm. 

Larger cooling power could be obtained by taking very large arrays of parallel NWs, as sketched in Fig.~\ref{FIG1a} for 15 NWs only. 
Let us take a packing density $\approx 20 \% $ for a 2d NWs array. This corresponds to a NW diameter $\approx 10$ nm and a NW interspacing 
of $\approx 40$ nm. $2\times 10^{5}$ parallel NWs would make an array of width $\approx 1$cm. It could be used for extracting 
$\approx 1$mW from the source side of the substrate, in a $1$ cm large and a few $\mu$m long area located near the source electrode. 
This estimated value of $\approx 1$mW is done assuming at a temperature $T \approx 75$K and a bias $\delta V = 1.3$mV. It could be 
increased by increasing  $T$ or $\delta V$, if VRH transport holds at larger temperatures and if the linear response theory remains valid 
at larger bias voltages. Additional results concerning the perspectives given by large arrays of 1d MOSFETs for energy 
harvesting and hot spot cooling can be found in Ref.~\cite{Bosisio2015}).

\section{Summary}
We have considered phonon assisted transport between localized states using a model characterized by the density of states $\nu(E)$ and the 
localization length $\xi(E)$ of the 1d Anderson model, going beyond theories where the energy dependence of $\nu$ and $\xi$ are neglected. 
We have studied the distributions of the heat currents characterizing locally the heat exchange between the substrate phonons and the 
NW electrons, as one shifts with a gate the NW conduction band. From those local heat currents, the structure of the percolative 
network connecting states of different energies $E_i$ and locations $x_i$ can be guessed. When the electrons are injected in the middle 
of the conduction band, the percolative path consists mainly of states with energy $E_i$ in an energy interval of width $\Delta_M$ around 
the band center. Even in this case, Fig.~\ref{2Dhistograms_IQi_T0p5} (a) shows us that the heat currents have larger fluctuations near the NW 
boundaries that in its bulk. This has to be related to the observation made in Ref.~\cite{Jiang2013} that the thermopower in the hopping regime 
is governed by the edges of the samples. When electrons are injected around the band edges, the boundary effects upon the local heat currents 
become much larger. It becomes possible to have mainly phonon absorption  at one NW boundary, and phonon emission at the other, instead of 
spreading phonon emission and absorption along the NW. In Refs.~\cite{Bosisio20142,Bosisio2015}, the thermoelectric effects of gated disordered 
NWs have been studied in the hopping regime. In this work, we have focused the study on the heat exchanges occurring between the NW electrons and 
the substrate phonons. The obtained results lead us to put forward arrays of 1d MOSFETs as tools for managing heat at submicron scales. Moreover, 
such tools make possible to control phonon absorption/emission by varying the gate and bias voltages.

%% The Appendices part is started with the command \appendix;
%% appendix sections are then done as normal sections
%% \appendix

%% \section{}
%% \label{}

%% If you have bibdatabase file and want bibtex to generate the
%% bibitems, please use
%%
\bibliographystyle{elsarticle-num} 
\bibliography{PAPER_BGFP}

\end{document}